\documentclass[12pt,preprint]{aastex}

\def\Msunh{\mbox{$h^{-1}\ M_\odot$}}

\shorttitle{The Halo Mass Function Redshift Dependence}
\shortauthors{Betancort-Rijo \& Montero-Dorta}

\begin{document}

\title{The Halo Mass Function Redshift Dependence}

\author{Juan E. Betancort-Rijo$^{1,2}$, Antonio D. Montero-Dorta$^{2,1}$}

\affil{$^{1}$Instituto de Astrof\'{i}sica de Canarias, V\'{i}a L\'{a}ctea s/n, 
 La Laguna, Tenerife, E38200, Spain}
\affil{$^{2}$Universidad de La Laguna, Departamento de Astrof\'{i}sica, Avda. Astrof\'{i}sico Fco. Sánchez s/n. ,E38200, La Laguna, Tenerife, Spain}
 
\email{jbetanco@iac.es}

\begin{abstract}
It has recently been shown that the correct halo mass function high mass behaviour may be obtained by an appropriate treatment of the all-mass-at-center problem. Here we show that the treatment of this problem leads to an accumulated mass fraction, $F$, which is not a universal function of the linear variance of the density contrast, but depends also on the shape of the power spectrum. For relevant power spectra $F$ is a universal function of two quantities: the linear variance and a quantity related to the local shape (at the given mass scale) of the power spectrum. This implies, for a given shape of the power spectrum, a redshift dependence of the mass function, which is quite important for a cold dark matter power spectrum. Our results are in good agreement with numerical simulations.  
           
\end{abstract}


\section{Introduction}
The dependence of the halo mass function on redshift is a theoretical issue of considerable interest. On the one hand it provides a test for our set of analytical tools for understanding the disipationless formation of structures out of Gaussian initial conditions. On the other hand, it may be an important ingredient in filling in the details of the standard model. For example, it may be particulary relevant to obtain the prescise reionization history of the universe (\citet{BL04}), since the number density of objects with mass $\sim10^{5-6}\Msunh$ as a function of redshift must be known with some accuracy to this purpose.\\
The question of the dependence of the mass function on redshift is a part of the larger question of the dependence on the power spectrum, $P(k)$. In fact, a change of redshift merely implies a change in the amplitude of $P(k)$. Formaly, the accumulated mass fraction, $F$, is, in principle, a function of mass and a functional with respect to $P(k)$: $F(m,P(k))$. However, the Press \& Schechter formalism (1974) leads to mass fraction depending on mass only through the $rms$ linear density contrast, $\sigma(m)$. In the excursion set formalism (\citet{PH90}, \citet{Bo91}) there exist the possibility of an additional dependence on $P(k)$, since, for top-hat filtering in real space, the relevant barrier penetration is not exactly a random walk process, and the way in which the actual process differs from a random walk at each scale depends on $P(k)$. However, this dependence is negligible, and the same applies when ellipsoidal dynamics is considered (\citet{ST99}). This implies that the mass function at any redshift may be obtained by rescaling $\sigma(m)$; that is, that $F$ is a universal function of $\sigma$.\\
However, evidence has been gathering from numerical simulation (\citet{Jen01}; \citet{War05}; \citet{Re06}) for an additional dependence of $F$ on $z$:

\begin{displaymath}
F(m,z)=\overline{F}(\sigma(m,z),z) \quad ; \quad \sigma(m,z)=D(z)\sigma(m)
\end{displaymath}

where $D(z)$ is the growth factor normalized to $1$ at present. Now, since for a given $\sigma$ value, $z$ is a function of $m$:

\begin{displaymath}
F(m,z)=\overline{F}(\sigma(m,z),z(m,\sigma))=F\prime(\sigma,m)
\end{displaymath}

So, the additional dependence of the mass fraction on $z$ reflects an additional (to that through $\sigma$) dependence on mass, which, in turn, reflects a dependence on $P(k)$, whose ``local'' shape changes with $m$.\\
The object of this work is to show how this dependence of the mass fraction on $z$, and, in general, on $P(k)$, may be determined using results from a recent work (\citet{BM06}), where the high mass behaviour of the mass function was explained.\\
We show here that the redshift dependence is just another aspect of the same effect, and that the mass fraction is a universal function of two quantities, $\sigma$, and a ``local''  spectral index that we represent by $c$.

\section{Framework}

We have shown in a recent work (\citet{BM06}) that the high mass behaviour of the cosmic mass function (at $z=0$) may be fully accounted for by simply adding to previous derivations  (\citet{ST99}) a consideration that has been, so far, overlooked. This consideration consists of the observation that, to assign a mass element to an object with mass $m$, it is more appropriate to impose the condition that it lies within a lagrangian distance $Q(m)$ ($Q(m)$ being its lagrangian radius) from the center of a colapsed object with mass $m$ than to demand the mass element to be at the center of one of these objects.\\
Using this criteria within the context of the spherical colapse approximation we found that the fraction of mass in objects with masses larger than $m$, $F(m)$, satisfies the relationship\\

 \begin{equation}
F(m)P_{2}(\delta_l|m\prime>m)=\frac{\exp\big(-\frac{\delta_{l}^{2}}{2\sigma(m)^{2}}\big)}{\sqrt{2\pi}\sigma(m)} \quad for \quad \delta_{l}>\delta_{c}
\label{eq:eq1}
\end{equation}   

where $P_{2}(\delta_l|m\prime>m)$ is the probability distribution for the value, $\delta_{l}$, of the linear density contrast within spheres of lagrangian radius $Q(m)$ centered at a randomly chosen mass element belonging to colapsed objects with mass larger than $m$. $\delta_{c}$ is the colapse linear density contrast in the spherical model, and $\sigma(m)$ is the $RMS$ linear density fluctuation on scale $m$. \\
In the referred work we obtained $P_{2}$ using an approximation that allows an analytical derivation. In this work we also use this approximation to most purposes, since it is rather accurate. However, with the main purpose of proving its accuracy explicity, we give here the exact expression for $P_{2}$ and briefly comment its relationship with that given in the reference.

\begin{equation}
P_{2}(\delta_{l}|m\prime>m)=\int_{m}^{\infty}G(m\prime,m)M\big(\delta_{l},m\prime,m\big) dm\prime 
\label{eq:eq2}
\end{equation}

where $G(m\prime,m)$ is the probability distribution for the mass, $m\prime$, of the object where a randomly chosen mass element belonging to objects with mass larger than $m$ is to be found. From the definition of $G(m\prime,m)$ and $F(m)$ it follows inmediatly that:

\begin{equation}
G(m\prime,m)=-\frac{1}{F(m)}\frac{dF(m\prime)}{dm\prime}
\label{eq:eq3}
\end{equation}

The quantity $M(\delta_{l},m\prime,m)$ represents the probability distribution of the linear density contrast at a sphere with radius $Q(m)$ centered at a randomly chosen mass element belonging to objects with mass $m\prime$. Through an straightforward extension of the arguments presented in the reference, we obtain for $M$:

\begin{equation}
M(\delta_{l},m\prime,m)=\frac{3}{\sqrt{2\Pi}}\int_{0}^{1}\frac{ \exp\bigg(\frac{-(\delta_{l}-\delta_{c}S(u))^2}{2\sigma\prime(u)^{2}}\bigg)}{\sigma\prime(u)}
u^{2}du
\label{eq:eq4}
\end{equation}

\begin{displaymath}
\sigma\prime(u)\equiv\sigma\prime(u,Q\prime,Q)=(1-S(u)^{2})^{\frac{1}{2}}\sigma(Q)
\end{displaymath}

where $S(u)$ is the dimensionless correlation between the linear density contrast on scale $Q(m\prime)$ and on scale $Q(m)$ at points lying at a lagrangian distance $uQ\prime$ from each other.

\begin{equation}
S(u)\equiv S(u,Q,Q\prime)=\frac{1}{2\Pi^{2}q}\frac{\int_{0}^{\infty}|\delta_{k}|^{2}W(Q\prime k)W(Qk)sin(kuQ\prime)kdk}{\sigma(Q)\sigma(Q\prime)}
\label{eq:eq5}
\end{equation}

\begin{displaymath}
W(x)=\frac{3}{x^{3}}(sin~x~-x~cos~x~) 
\end{displaymath}

where $|\delta_{k}|^{2}$ is the power spectrum of the linear density contrast.\\
Inserting expression (\ref{eq:eq3}) and (\ref{eq:eq4}) into expression (\ref{eq:eq2}) and this last expression into (\ref{eq:eq3}) we obtain an integro-differential equation for $F(m)$, which, integrating by parts, may be transformed into an integral equation:

\begin{equation}  F(m)=\frac{1}{M(\delta_{l},m,m)}\bigg[\frac{exp\bigg(-\frac{\delta_{l}^{2}}{2\sigma(m)^{2}}\bigg)}{\sqrt{2\Pi}\sigma(m)}+\int_{m}^{\infty}F(m\prime)\frac{dM(\delta_{l},m\prime,m)}{dm\prime}dm\prime\bigg]
\label{eq:eq6}
\end{equation}

However, in practice, it may be more convenient to use the integro-differential equation. Equation (\ref{eq:eq6}) (or its integro-differential counterpart) is valid for any $\delta_{l}\geq\delta_{c}$ (it comes from expression (\ref{eq:eq1})), so, for any $\delta_{l}\geq\delta_{c}$ the right hand side must be independent of $\delta_{l}$.\\
In the referred work we implicity used $M(\delta_{l},m,m)$ (that we called there $P_{1}(\delta_l|\delta_l(q=0)=\delta_c,Q)$) in expression (\ref{eq:eq2}) instead of $M(\delta_{l},m\prime,m)$. This implies that mass elements within objects with mass larger than $m$ were assumed to lie within distance $Q(m)$ from the center of the object. This is not as bad as assuming them all to be at the very center, but it is not as good as letting them be within distance $Q(m\prime)$ from the center of the object, with mass $m\prime$, to which they actually belong. However, in the relevant high mass region, this assumption does not involve much inaccuracy, since most objects larger than $Q(m)$ are just $Q(m)$ plus a bit.\\
Under this assumption, expression (\ref{eq:eq2}) may be transformed from an integral over mass into the integral of expression (\ref{eq:eq4}) (with $w$ in the place of $\delta_{c}$) over all $w\geq\delta_{c}$, which may be integrated analytically to give, when inserted into (\ref{eq:eq1}):

\begin{equation}
F(m)=\frac{erfc\bigg(\frac{\delta_{c}}{\sqrt{2}\sigma(m)}\bigg)}{V(m)}  \label{eq:eq7}
\end{equation}
 
\begin{displaymath}
V(m)\equiv3\int_{0}^{1}erfc\bigg(\frac{\delta_{c}}{\sqrt{2}\sigma(m)}\bigg(\frac{1-\overline{S}(u)}{1+\overline{S}(u)}\bigg)^{\frac{1}{2}}\bigg)u^{2}du
\end{displaymath}

\begin{displaymath}
\overline{S}(u)\equiv S(u,Q,Q) \quad;\quad Q=Q(m)
\end{displaymath}

where $S$ is given by (\ref{eq:eq5}). To simplify the evaluation of $V(m)$, the following accurate aproximation for $S$ may be used for some power spectra:

\begin{equation}
S(u,Q,Q)\sim e^{-cu^{2}} \quad;\quad c\equiv c(Q)=Ln(S(1,Q,Q))
\label{eq:eq8}
\end{equation}

$c(Q)$ is a coefficient depending on $Q$ but not on $u$, which may be obtained evaluating $S(1,Q,Q)$ as indicated in expression (\ref{eq:eq5}).

\section{The redshift dependence}

As mentioned in the introduction, the accumulated mass function, $F(m)$,  derived within the excursion set formalism depends on mass only through $\sigma(m)$. This is referred to as the universality of the mass fractions, both the accumulated, $F(\sigma)$, and the differential, $f(\sigma)$ (the derivative of $F(\sigma)$ with respect to $Ln\sigma$). That is, for a given value of $\sigma$, the mass fractions are independent of the power spectrum, and, in particular, are independent of redshift, since changing the redshift merely implies a change of the amplitude of the power spectrum. However, the mass fraction obtained within the present framework (i.e. with an appropriate treatment of the all-mass-at-center problem) do not show this universality. From approximation (\ref{eq:eq7}), we see that $F(m)$ does not depend on $m$ only through $\sigma(m)$. $V(m)$ depends on $m$ through $\sigma$ and through $\overline{S}(u)$, whose functional form depends on $m$ (through $Q(m)$) and, in general, on the whole functional form of the power spectrum. If the power spectrum takes the form:

 \begin{displaymath}
 P(k)=A(k)k^{n_{eff}(k)} 
 \end{displaymath}
 
with $n_{eff}$, $A$ slowly varying functions of k, then approximation (\ref{eq:eq8}) holds and $V$ depends on $m$ only through $\sigma(m)$ and $c(m)$. Therefore, for these power spectra, the accumulated mass fraction is a universal function of $\sigma$, $c$ \quad:

\begin{equation}
F_{0}(\sigma,c)=erfc\bigg(\frac{\delta_{c}}{\sqrt{2}\sigma(m)}\bigg)\bigg(3\int_{0}^{1}erfc\bigg(\frac{\delta_{c}}{\sqrt{2}\sigma(m)}\bigg(\frac{1-\exp(-c(m)u^{2})}{1+\exp(-c(m)u^{2})}\bigg)^{\frac{1}{2}}\bigg)u^{2}du\bigg)^{-1}
\label{eq:eq9} 
\end{equation}

giving a differential mass function:

\begin{equation}
n_{0}(m)=-\frac{\rho_{b}}{m}\frac{dF_{0}}{dm}(\sigma(m),c(m))
\label{eq:eq10}
\end{equation}

$c(m)$ is a sort of ``local'' spectral index which depends on the ``local'' form of the power spectra, but not on its amplitude. In a coarse approximation we have:

\begin{equation}
c(m)\simeq\frac{n_{eff}+3}{4} \quad;\quad k(m)\sim\frac{2\pi}{Q(m)}
\label{eq:eq11}
\end{equation}

For a CDM power spectrum with $\Omega_{m}=0.3$, $h=0.7$ (\citet{BBKS86}) $c(m)$ may be accurately obtained with the following fit:

\begin{equation}
c(m)=2.347\cdot10^{-6}-1.678\cdot10^{-4}Ln(m)+4.888\cdot10^{-3}(Ln(m))^{2}-5.909\cdot10^{-2}(Ln(m))^{3}+0.367(Ln(m))^{4}
\label{eq:eq12}
\end{equation}

Given expression (\ref{eq:eq9}) and (\ref{eq:eq10}) for the present cosmic mass function, it is inmediate to obtain the mass function at redshift $z$: \quad $\sigma(m)$ must be replaced by its rescaled value $\sigma(m,z)$ while $c(m)$ retain its value. Now, the mass corresponding to a given $\sigma$ value is smaller at redshift $z$ than at present, so that, being $c(m)$ a monotonicaly increasing function, the value of $c$ corresponding at redshift $z$ to a given value of $\sigma$ is smaller than at present. But, since, at fixed $\sigma$, $F$ is monotonically increasing with $c$, this implies that for a given $\sigma$ the value of $F$ decreases with increasing redshift. Therefore, we see that, for a fixed form of the power spectrum, the dependence of $F$ on $c$ translates into a redshift dependence for the dependence of $F_{0}$ on $\sigma$:

\begin{equation}
F(\sigma,z)=F_{0}(\sigma,c(m(\sigma,z)))
\label{eq:eq13} 
\end{equation}

where $m(\sigma,z)$ is obtained solving the equation $D(z)\sigma(m)=\sigma$.\\
All we have said so far concerns the mass function obtained within the spherical colapse model. In a forthcoming work we will show how to combine our formalism with ellipsoidal dynamics. However, the dependence of $F$ on $c$ arises from the appropriate treatment of the all-mass-at-center problem that we know is relevant only in the high mass region $(\sigma(m)\lesssim1)$ where the spherical colapse model is a good approximation. Therefore, if we combine the $c$ dependence of $F$ obtained within the spherical model with the $\sigma$ dependence at $z=0$ obtained from numerical simulations, we may obtained a good approximation to the redshift dependence of the mass function.\\
Let us see how this can be done. The excursion set formalism when combined with ellipsoidal dynamics leads to Sheth \& Tormen (\citet{ST99}) analytic mass function (with $a=1$). In \citet{BM06} we conjectured that an appropriate treatment of the all-mass-at-center problem within the context of ellipsoidal dynamics will increase the S-T mass fraction by approximately the same factor as that enhancing the normalized P-S mass fraction when the treatment of that problem is implemented within the spherical model. Expression (\ref{eq:eq7}) tells us that this factor is simply $V^{-1}$. So, within this approximation we have for the present mass fraction:

\begin{equation}
F(m)=F_{S-T}(\sigma(m))\bigg(V(\sigma(m),c(m))\bigg)^{-1}\quad;\quad F_{S-T}(\sigma)=\int_{-infty}^{Ln\sigma}f_{S-T}(\sigma\prime)dLn\sigma\prime
\label{eq:eq14}
\end{equation}

where $F_{S-T}$ stands for the S \& T theoretical prediction for the mass fraction (i.e. that corresponding to their fitting formula with $a=1$) and $f_{S-T}$ for its corresponding differential mass fraction. We have seen (Betancort \& Montero, 2006) that $F(m)$ is in good agreement with the fits to numerical simulations of the present mass function.\\
Expression (\ref{eq:eq14}) implies that $F(m)$ does not depend on $m$ only through $\sigma$. However, at present, $\sigma$ is a unique function of $m$, so we may find a function of $\sigma$, $\overline{F}$, such that $\overline{F}(\sigma(m))=F(m)$. Using (\ref{eq:eq14}) we have for $\overline{F}$:

\begin{equation}
\overline{F}(\sigma)=F_{S-T}(\sigma)\bigg(V(\sigma,c(\overline{m}(\sigma)))\bigg)^{-1} 
\label{eq:eq15}
\end{equation}

where $\overline{m}(\sigma)$ is the inverse function of $\sigma(m)$ (at present). Now, if, working under the wrong assumption that the mass fraction is only a function of $\sigma$, we rescaled $\sigma$ with redshift, we would not obtain the correct mass function at that redshift, but something whose relationship with the actual mass function, which follows inmediately from (\ref{eq:eq14}) and (\ref{eq:eq15}), is given by:

\begin{equation}
F(m,z)=\overline{F}(\sigma(m,z))g(m,z) \quad;\quad g(m,z)=\frac{V(\sigma(m,z),c(\overline{m}(\sigma(m,z))))}{V(\sigma(m,z),c(m))}
\label{eq:eq16}
\end{equation}

\begin{displaymath}
\sigma(m,z)=D(z)\sigma(m)  
\end{displaymath}

By simply rescaling $\overline{F}(\sigma)$ we are implicitly evaluating $V$ not at the nominal mass, $m$, but at $\overline{m}$ (with $\sigma(\overline{m})\equiv\sigma(\overline{m},0)=\sigma(m,z)$). So, to obtain the correct mass function, the rescaled $\overline{F}$ must be multiplied by the ratio between the values of $V^{-1}$ at $m$ and at $\overline{m}$. For the mass functions themselves we have from (\ref{eq:eq16}) and (\ref{eq:eq10}):

\begin{equation}
n_{0}(m,z)=-\frac{\varrho_{b}}{m}\overline{f}(\sigma(m,z))\bigg[\frac{dLn(\sigma(m,z))}{dm}g(m,z)+\overline{F}(\sigma(m,z))\frac{dg(m,z)}{dm}\bigg]
\label{eq:eq17}
\end{equation}

\begin{displaymath}
\sigma(\overline{m})=D(z)\sigma(m)
\end{displaymath}

where $\overline{f}(\sigma)$ is the differential mass fraction corresponding to $\overline{F}$. Any fit to numerical simulations may be used for $\overline{f}(\sigma)$. We shall use S \& T fit ($a=0.707$).\\
In figure (1) we show the ratio between the mass function given by (\ref{eq:eq17}) and the rescaled S \& T fit to the mass function ($a=0.707$). For $z=10$ the ratio is close to one except for the largest masses. For $z=20$ the actual mass function is a 35\% smaller than the rescaled one at $m=10^{5}\Msunh$, the difference being a factor $4$ for $m=10^{9}\Msunh$. For $z=30$ the difference is already a factor $2.7$ for $m=10^{5}\Msunh$ and is larger than a factor $6$ for $m=10^{8}\Msunh$. This figure is in good agreement with the numerical results presented by Reed et al (2006) in their figure A3.\\
Finally, to check that our results are not strongly dependent on the fact that we have used approximation (\ref{eq:eq9}), we recalculate the factor $g(m,z)$ using expression (\ref{eq:eq6}), which is exact within the spherical model. We solve the equation iteratively using for $F(m\prime)$ in the integrand approximation (\ref{eq:eq7}). We have also used for  $F(m\prime)$ that corresponding to the normalized P-S mass function (i.e. expression (\ref{eq:eq7}) with $V=1$), obtaining very similar results. From this we may infer that the values of $n$ that we have obtained with expression (\ref{eq:eq6}), $n_{1}$, has at most a 2 \% error for all masses and redshift considered. The ratio between $n_{1}$ and the value of $n$ given by expression (\ref{eq:eq10}), $n_{0}$, is shown in figure (2). We see that this ratio is larger than one, thereby increasing the redshift dependence of the mass function, but the logarithmic correction is small (i.e. $Ln(n_{0})/Ln(\frac{n_{1}}{n_{0}})\gg1$). We may therefore conclude that the results presented here for the redshift dependence of the mass function will remain basicaly unchanged in more rigorous analysis of the problem.

\clearpage

\begin{figure}
\includegraphics[scale=0.65,angle=90]{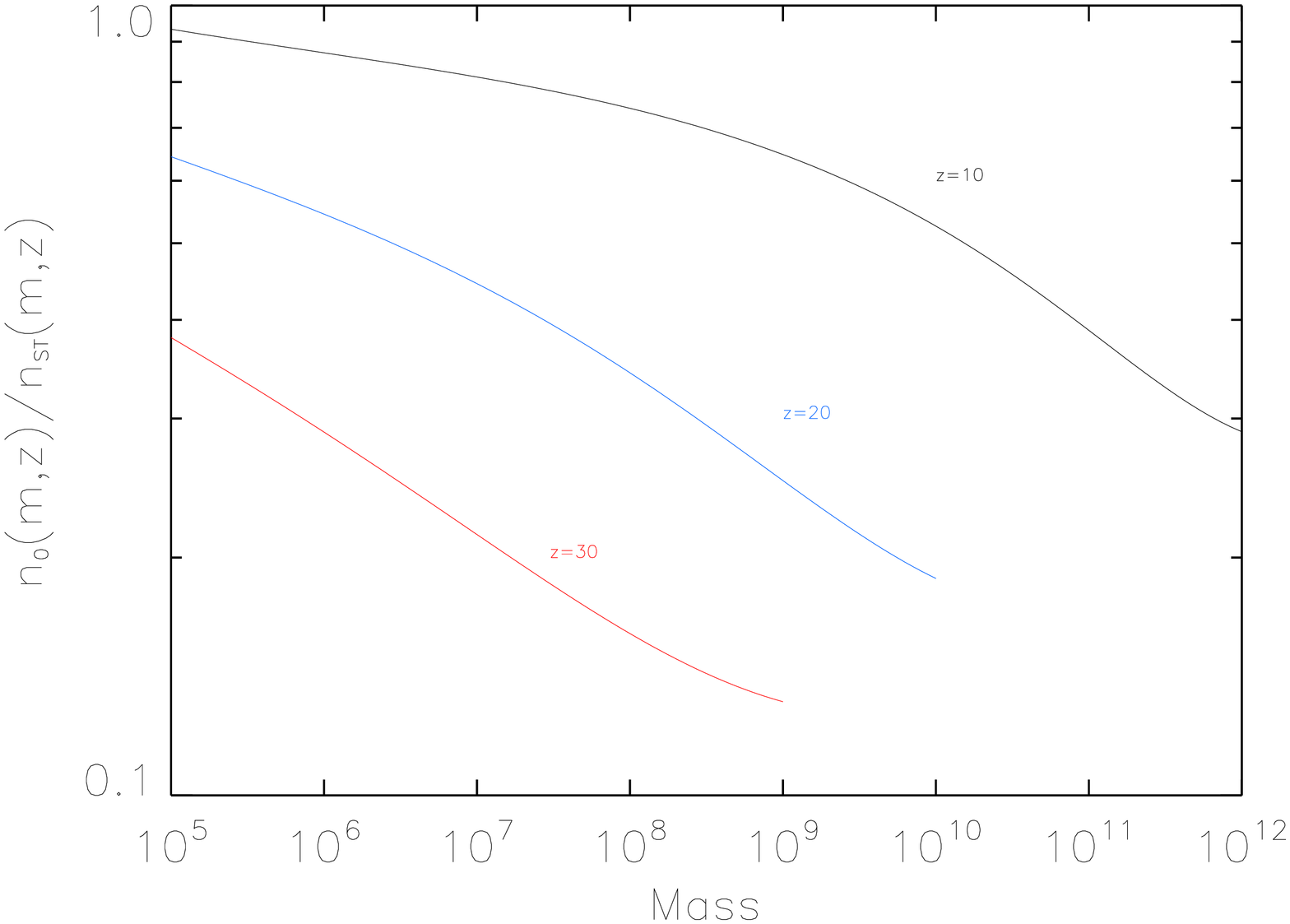}
\caption{{The mass function given by expression (\ref{eq:eq17}) normalized to the rescaled S \& T mass function ($a=0.707$) at three different redshifts.}}
\end{figure}

\clearpage

\begin{figure}
\includegraphics[scale=0.65,angle=90]{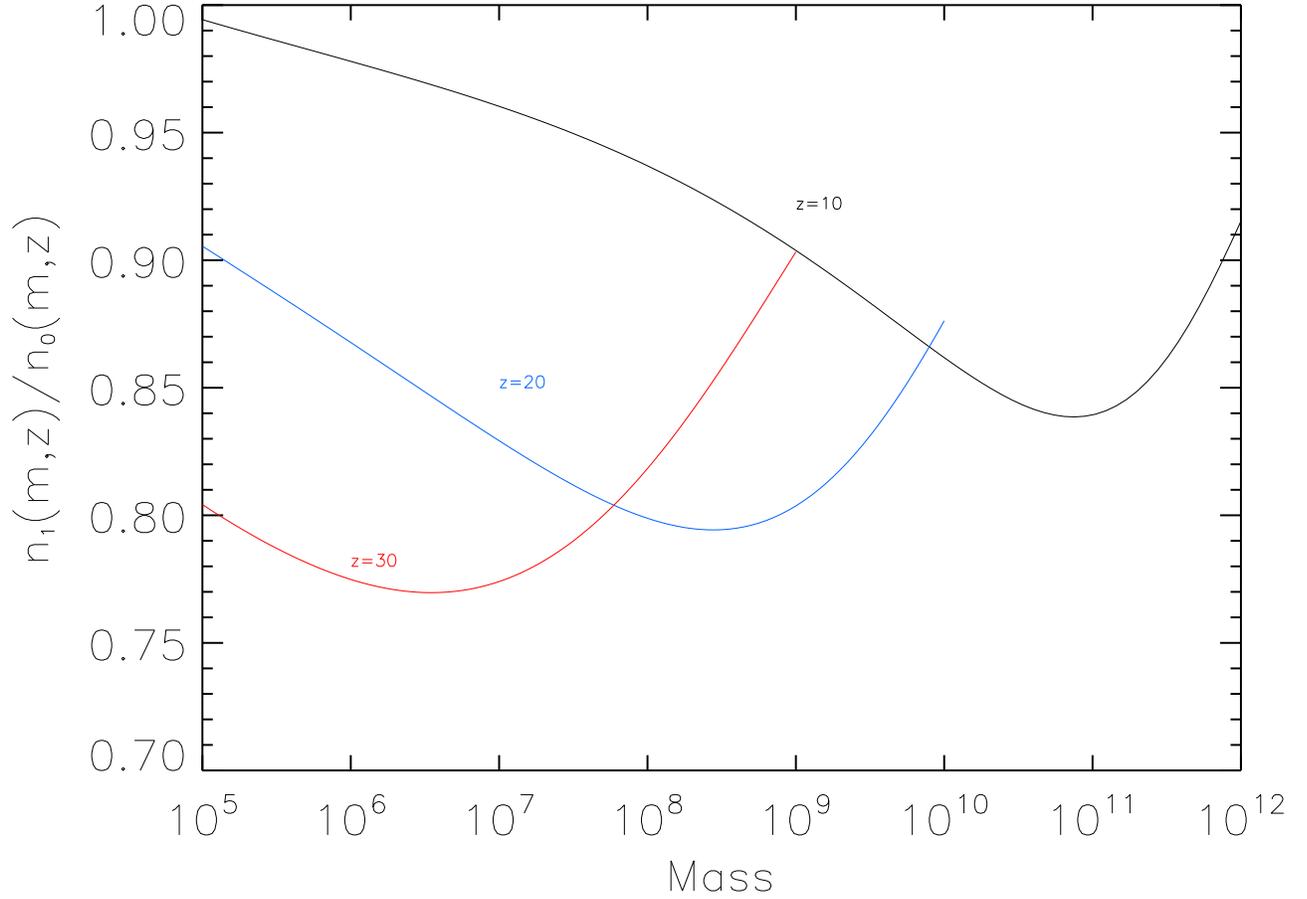}
\caption{{Ratio between the mass functions obtained with the exact expression (expression (\ref{eq:eq6})), $n_{1}$, and that obtained with approximation (\ref{eq:eq17}), $n_{0}$, at three different redshifts.}}
\end{figure}

\end{document}